# Three-dimensional topological semimetal phase in layered TaNiTe$_5$ probed by de Haas-van Alphen effect

Zheng Chen[1,2,*], Min Wu[1,*], Yong Zhang[1,3], Jinglei Zhang[1], Yong Nie[1,2], Yaru He[1,2], Yongliang Qin[1], Yuyan Han[1], Chuanying Xi[1], Shuaiqi Ma[1,3], Xucai Kan[3], Jianhui Zhou[1], Xiaoping Yang[1], Xiangde Zhu[1,‡], Wei Ning[1‡], and Mingliang Tian[1,3,4‡]

[1]*Anhui Province Key Laboratory of Condensed Matter Physics at Extreme Conditions, High Magnetic Field Laboratory, Chinese Academy of Sciences, Hefei 230031, Anhui, China*

[2]*Department of Physics, University of Science and Technology of China, Hefei 230026, China*

[3]*School of Physics and Materials Sciences, Anhui University, Heifei 230601, China*

[4]*Collaborative Innovation Center of Advanced Microstructures, Nanjing University, Nanjing 210093, China*

[*] Those authors contribute equally to this work.

[‡] To whom correspondence should be addressed. E-mail: xdzhu@hfml.ac.cn (X.Z.), ningwei@hmfl.ac.cn (W.N.), tianml@hmfl.ac.cn (M.T.)

Layered three-dimensional (3D) topological semimetals have attracted intensively attention due to the exotic phenomena and abundantly tunable properties. Here we report the experimental evidence for the 3D topological semimetal phase in layered material $TaNiTe_5$ single crystals through quantum oscillations. Strong quantum oscillations have been observed with diamagnetism background in $TaNiTe_5$. By analyzing the de Haas-van Alphen oscillations, multi-periodic oscillations were extracted, in content with magnetotransport measurements. Moreover, nontrivial $\pi$ Berry phase with 3D Fermi surface is identified, indicating the topologically nontrivial feature in $TaNiTe_5$. Additionally, we demonstrated the thin-layer of $TaNiTe_5$ crystals is highly feasible by the mechanical exfoliation, which offers a platform to explore exotic properties in low dimensional topological semimetal and paves the way for potential applications in nanodevices.

Three-dimensional (3D) topological semimetals, Dirac and Weyl semimetals, are quantum matter that possess low-energy quasiparticle excitations (massless fermions), and have attracted extensive interest in the past few years [1-4]. Specially, the layered 3D topological materials with van der Waals (vdW) coupling have a fertilely tunable properties by quantum confinement effects or vdW heterojunctions [5-8]. Up to now, only a few vdW materials that have been verified to host topologically nontrivial nature, such as $XTe_2$ (X=W, Mo) [9-11], $ZrTe_5$ [12,13], and $TaIrTe_4$ [14,15].

The layered ternary transition metal chalcogenide $TaNiTe_5$ was firstly synthesized and reported to be antiferromagnetic and highly conductive metal in 1989 [16]. However, the investigation on its electric structure remained unexplored. Recently, the vdW material $TaNiTe_5$ was theoretically predicted to be topologically nontrivial semimetal [17], but no experimental evidence has been reported yet.

In this work, we systematically study the topological properties by analyzing the quantum oscillations in $TaNiTe_5$ single crystals. The intrinsic diamagnetism nature in this layered material was established by the magnetization measurements. Strong de Haas-van Alphen (dHvA) oscillations were observed at low temperatures in the magnetic torque measurements. From the fast Fourier transform (FFT) analysis of the dHvA oscillation components, multi-periodic oscillations were obtained, which was verified by the Shubnikov-de Haas (SdH) oscillations obtained from magnetotransport experiment in high magnetic field (30 T). Furthermore, nontrivial Berry phase and 3D Fermi surface with small effective mass were identified. Our experimental results provide a strong evidence for the 3D topological semimetal phase in layered material

TaNiTe$_5$ single crystals.

The high quality TaNiTe$_5$ single crystals were synthesized via Te flux method. The molar ratio of elements was Ta (powder): Ni (wire): Te (ingot) = 1:1:10. The mixture agents in the alumina crucible with an inverted crucible above were sealed into a quartz tube. The tube was heated to 700 °C at 30 °C/$h$ and held for 4 days, then slowly cooled to 500 °C in a week. The flux was removed by quick centrifugation.

The obtained crystals show rectangular shape with the length up to several millimeters, as shown in the inset of Fig. 1(b). The crystal structure of TaNiTe$_5$ has an orthorhombic layered structure with a = 3.659 Å, b = 13.122 Å, c = 15.111 Å and space group *Cmcm* (No. 63) [16]. In this layered structure, the trigonal NiTe$_2$ chains arrange along the crystallographic $a$-axis and form a two-dimensional layer via linking chains of Ta atoms along the $c$-axis. The sandwiched layer stacks along the $b$-axis into a crystal, as shown in Fig. 1(a). The X-ray diffraction (XRD) was used to determine the quality and structure of the as-grown TaNiTe$_5$ single crystals. As depicted in Fig. 1(b), all the peaks can be indexed as (0 $L$ 0) reflections of TaNiTe$_5$ crystals and no trace of impurity phase is detected, which indicates that the largest nature-grown surface of TaNiTe$_5$ crystals belongs to $a$-$c$ plane and high crystallinity of the crystals, respectively. In our experiment, the magnetization was measured on superconducting quantum interference device (SQUID, Quantum Design, Inc.). The magnetic torque measurements were carried out on Oxford cryostat equipped with a 12 T superconducting magnet, using capacitive cantilever method. The magnetotransport at high magnetic field were performed using standard ac lock-in techniques with a water-cooled magnet (30 T) at the Chinese High Magnetic Field Laboratory (CHMFL) in Hefei.

Fig. 1(c) shows the temperature dependence of resistivity measured with the current applied along $a$-axis at zero field (left side). A typical metallic behavior with RRR = 12 was observed, that is well consistent with previous work [16], where its susceptibility obeys the Curie-Weiss law in temperature range from 2.5 to 200 K. On the contrary, the negative magnetic susceptibility reveals a diamagnetism feature in our samples (right side), as indicated in Fig. 1(c). Generally, impurities induced long-range spin ordering can easily smear the inherent diamagnetism [18]. Thus, the observation of diamagnetism indicates the high quality of single crystals studied in this work. This negative magnetic susceptibility was also observed in other nonmagnetic topological semimetals, such as Weyl semimetal TaAs [19,20], Dirac semimetal $Cd_3As_2$ [21], $PdTe_2$ [22], and nodal-line semimetals ZrSiS family [23,24].

Fig. 1(d) presents the dHvA effect, the quantum oscillations in magnetization, which was superposed on the diamagnetism background measured by SQUID with magnetic field parallel to $b$-axis. The presence of dHvA oscillations can be traced down to a field of ~4 T at $T = 1.8$ K, as shown in the inset of Fig. 1(d). With increasing temperature, the amplitude of oscillations decreases rapidly and ultimately vanishes at 15 K, as shown in Fig. 1(d). Noted that, the SdH oscillations are not observed with magnetic field up to 12 T, as shown in the inset of Fig. 1(c), which may due to the carrier mobility degraded by the electrical contacts. The Hall measurements (inset of Fig. 1(c)) indicate the hole carriers dominant transport with the carrier density $n_h =$

$3.5 \times 10^{19} \text{cm}^{-3}$ at $T = 2$ K.

To get a comprehensive insight into the quantum oscillations and determine the nontrivial nature of band structure in TaNiTe$_5$, we carried out the magnetic torque measurements with field up to 12 T, which is a useful tool for exploring the electron states in solids [25,26]. Fig. 2(a) plots the magnetic field dependence of magnetic torque $\tau$ at different temperatures with $B \parallel b$-axis. As it clearly shows, the presence of pronounced dHvA oscillations with multiple frequencies can be traced down to a field of $\sim 4$ T at $T = 2$ K (inset of Fig. 2(a)) and survived up to 20 K. The oscillation components $\Delta\tau$ obtained after subtracting the smooth diamagnetism background from $\tau$, are plotted in Fig. 2(b) as a function of the inversed field $1/B$. The amplitude of oscillations decreases rapidly with the increase of temperature. By applying the fast Fourier transform (FFT), four fundamental oscillation frequencies are identified, $F_\alpha = 60$ T, $F_\beta = 184$ T, $F_\gamma = 844$ T, and $F_\eta = 1021$ T, as shown in Fig. 2(c). According to the Onsager relation $F = \frac{\Phi_0}{2\pi^2} A_F$, the cross-sectional areas of Fermi surface normal to the field is about $A_F = 0.006$ Å$^{-2}$, $0.018$ Å$^{-2}$, $0.081$ Å$^{-2}$, and $0.098$ Å$^{-2}$ with the Fermi wave vector estimated to be $k_F = 0.044$ Å$^{-1}$, $0.076$ Å$^{-1}$, $0.161$ Å$^{-1}$ and $0.177$ Å$^{-1}$ for these four bands (assuming a circular cross section. The detailed geometrical shape of Fermi surface will be discussed below), respectively. Due to the signals for $\gamma$ and $\eta$ bands being very weak, we could not give an accurate information about these two band. Therefore, we mainly focus on the $\alpha$ and $\beta$ bands for analyzing.

In general, the magnetic torque is proportional to magnetization $\tau \propto \mathbf{M} \times \mathbf{B}$, which indicates that the oscillation amplitude of $\Delta\tau$ in a 3D system can also be described by the Lifshitz-Kosevich (LK) formula with the Berry phase included [27,28]:

$$\Delta\tau \propto -B^{1/2} R_T R_D \sin\left[2\pi\left(\frac{F}{B} - \gamma - \delta\right)\right],$$

where $R_T = \frac{2\pi^2 k_B T/\hbar\omega_c}{\sinh(2\pi^2 k_B T/\hbar\omega_c)}$, $R_D = \exp\left(-2\pi^2 k_B T_D/\hbar\omega_c\right)$ are the temperature-damping and field-damping factors, respectively. $k_B$ is the Boltzmann constant. $T_D$ is the Dingle temperature. $\hbar$ is the reduced Plank constant and the cyclotron frequency $\omega_c = eB/m^*$, with $m^*$ the effective cyclotron mass of electron. The oscillation components are proportional to the sine term with $\gamma = \frac{1}{2} - \beta$, where $2\pi\beta$ is the $\pi$ Berry phase and $2\pi\delta$ is the additional phase shift with $\delta$ determined by the dimensionality of Fermi surface and changed from 0 for 2D to $\pm 1/8$ for 3D (+for hole and − for electron bands). In Fig. 2(d), we plotted the temperature dependence of normalized oscillation amplitudes at $1/B = 0.1135$ T$^{-1}$ and $0.0956$ T$^{-1}$ for $\alpha$ and $\beta$ bands. The solid lines are the best fit to the experimental data, which generate small effective cyclotron mass $m_\alpha^* = 0.19\ m_0$ and $m_\beta^* = 0.22\ m_0$ for $\alpha$ (top panel) and $\beta$ (bottom panel) bands, respectively. Correspondingly, the Dingle temperature $T_D$ estimated from the slope of the Dingle plot is $T_D^\alpha = 17.3$ K and $T_D^\beta = 14.4$ K for $\alpha$ and $\beta$ pockets, respectively. Using the fitted Dingle temperature, the quantum relaxation time $\tau_Q = \frac{\hbar}{2\pi k_B T_D}$ and the quantum mobility $\mu_Q = \frac{e\tau_Q}{m^*}$ were calculated, which were shown in Table 1. What should also be point out is that the obtained small effective cyclotron mass could reasonably explain the diamagnetic susceptibility (Fig. 1(d)) according to the Landau diamagnetism theory, $\chi = \chi_p\left[1 - \frac{1}{3}\left(\frac{m_e}{m^*}\right)^2\right]$ with $\chi_p$ paramagnetic susceptibility [29].

The nontrivial π Berry phase is a key feature for Dirac fermions with linear dispersion and has been widely reported in various topological materials through the quantum oscillations spectra [30-32]. The Landau level (LL) fan diagram, that is the LL index $n$ versus $1/B$, is plotted in Fig. 2 (e). Here, the oscillation peaks and valleys are assigned as $n+\frac{1}{4}$ and $n-\frac{1}{4}$ [21,33], respectively. According to the Lifshitz-Onsager quantization rule $\frac{F}{B} = n + \gamma + \delta$, the LL index $n$ is linearly dependent on $1/B$. As shown in Fig. 2 (e), non-zero intercept extracted by the linear extrapolation to the $n$ axis is $-0.364$ and $0.349$ for $\alpha$ and $\beta$ band, respectively. Note that the intercept diverges from the expected value $\pm 1/8$ may due to the anisotropic ellipsoidal character of the Fermi surface, and the coexistence of parabolic dispersion around the Fermi level [17]. Finally, the Berry phase is calculated to be $(-0.73 \pm 0.25)\pi$ and $(0.70 \pm 0.25)\pi$ for $\alpha$ and $\beta$ bands, respectively. The nontrivial Berry phase for these two bands provides a strong evidence for the existence of 3D Dirac fermions in TaNiTe$_5$ crystals.

To exactly construct the morphology of Fermi surface in TaNiTe$_5$, we carried out the angle dependent quantum oscillation measurements. The strong dHvA oscillations in the angle range from $0°$ to $90°$ at $T = 2$ K is plotted in Fig. 3(a), where the $\theta$ is defined as the angle between the direction of field $B$ and the $b$-axis. The oscillation amplitude is suppressed gradually while the titled angle $\theta$ increases, as shown in Fig. 3(b). As we can see, multi-periodic oscillation patterns were unambiguously revealed at different angles, demonstrating 3D character of the Fermi surface in TaNiTe$_5$. To illustrate the evolution of the 3D Fermi surface, we present the corresponding FFT spectra for each titled angle $\theta$ in Fig. 3(c). With increasing angle $\theta$, the oscillation frequency for $\alpha$ band is almost unchanged (grey solid line), indicating a nearly spherical Fermi surface. However, varying $\theta$ has very strong effect on $\beta$ band, which displays a successive increase (except for $\theta = 90°$) in the oscillation period (grey dashed line). Moreover, the 2D model and standard 3D ellipsoid model of Fermi surface has been used to fit the oscillation frequency $F_\beta$ and $F_\gamma$ and as function of $\theta$ as function of $\theta$, and the results were shown in Fig. 3(d). Apparently, the experimental data can be well fitted with the 3D ellipsoid model (red solid line), implying that the $\beta$ band has an anisotropic 3D Fermi surface. Simultaneously, the $\gamma$ and $\eta$ bands are also exhibit 3D ellipsoidal Fermi surface, as shown in Figs. 3(c) and 3(d).

To further confirm the inherently multiple Fermi pockets in TaNiTe$_5$ crystals, we performed the magnetotransport measurements with the field up to 30 T. Fig. 4(a) shows the longitudinal magnetoresistance (MR), defined as MR(%) = $[R(B) - R(0)]/R(0) \times 100\%$, measured at different temperatures with the magnetic field along $b$-axis. The non-saturating linear MR at high field rang with multi-periodic SdH oscillations is revealed. This linear MR behavior has also been widely reported in topological semimetals, however, the origin remains controversy [34-36]. Furthermore, three oscillation frequencies are derived from the FFT spectra, $F'_\alpha = 59$ T, $F'_\beta = 180$ T, and $F'_\gamma = 770$ T, as shown in Fig. 4(b). The obtained results are in agreement with these extracted from the dHvA oscillations, expect for the tiny disparity in the values and the highest frequency unrevealed. Although the oscillations for both cases are attributed to the same physical mechanism, the fact that the SdH oscillations stem from the oscillations in the scattering rate of carriers while the dHvA oscillations originate from the oscillations in free energy reasonably explains the discrepancies in the results obtain by these two different methods.

In conclusion, we have carried out comprehensive studies to demonstrate the existence of a three-dimensional topological semimetal phase in layered TaNiTe$_5$ crystals. Multi-periodic dHvA oscillations of magnetic torque were found at low temperatures, which was further confirmed by the magnetotransport measurements. By carefully analyzing the dHvA oscillations, the nontrivial π Berry phase with 3D Fermi surface was revealed, which presents a clear evidence of the presence of 3D Dirac fermions in TaNiTe$_5$. Due to the vdW coupling between the interlayer, the thin layer of TaNiTe$_5$ (See Supplementary Material) is highly feasible by the mechanical exfoliation, which paves the way for potential applications in nanodevices.

Recently, we are aware of the study on the quantum oscillation measurements by C. Q. Xu [37].

**Acknowledgments**

This work was supported by the National Key Research and Development Program of China, Grant No. 2016 YFA0401003, the Natural Science Foundation of China (Grants No. U19A2093, 11774353, 11574320, 11374302, No. U1432251, and U1732274); Collaborative Innovation Program of Hefei Science Center, CAS (Grant No. 2019HSC-CIP 001), the Innovative Program of Development Foundation of Hefei Center for Physical Science and Technology (Grant No. 2018CXFX002).

# Figure Captions

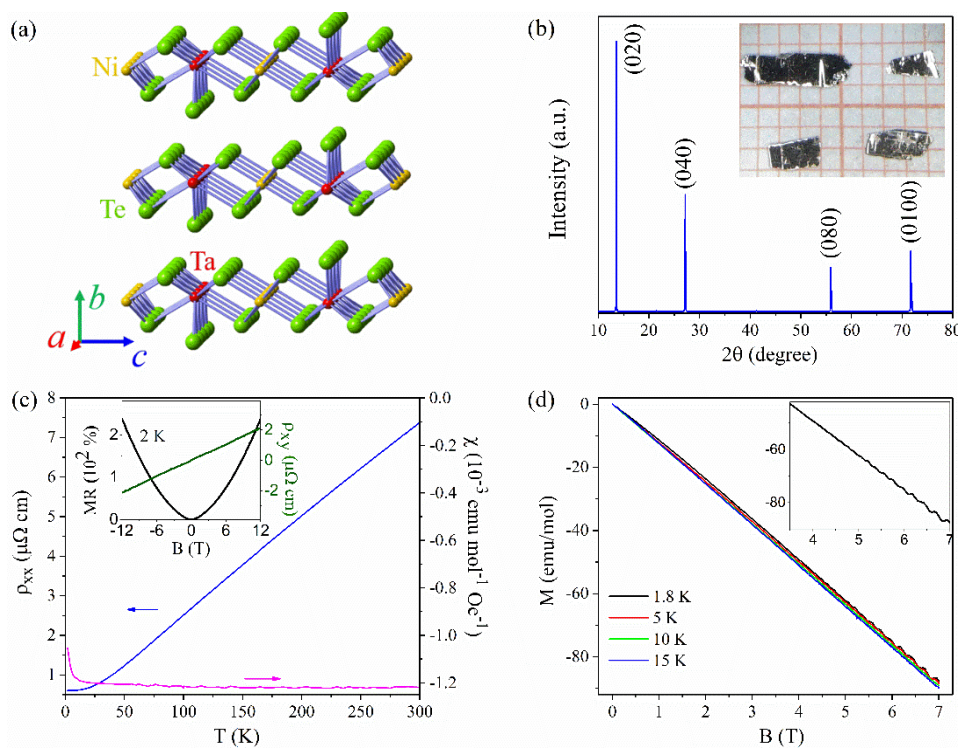

FIG. 1. (Color online) (a) Crystal structure of TaNiTe$_5$ single crystals with space group *Cmcm* (No. 63). (b) The large-grown surface is (010) surface that revealed by XRD pattern. Inset: Optical micrograph of the grown TaNiTe$_5$ single crystals. (c) Resistivity ($B = 0$) and magnetic susceptibility ($B = 0.05$ T) versus temperature of TaNiTe$_5$. Inset: The longitudinal MR (left axis) and Hall resistivity (right axis) as function of applied field $B$ at $T = 2$ K. (d) dHvA Oscillations of magnetization as function of $B$ at various temperatures with $B \parallel b$-axis. Inset: the oscillations can be traced down to a field of ~4 T at $T = 1.8$ K.

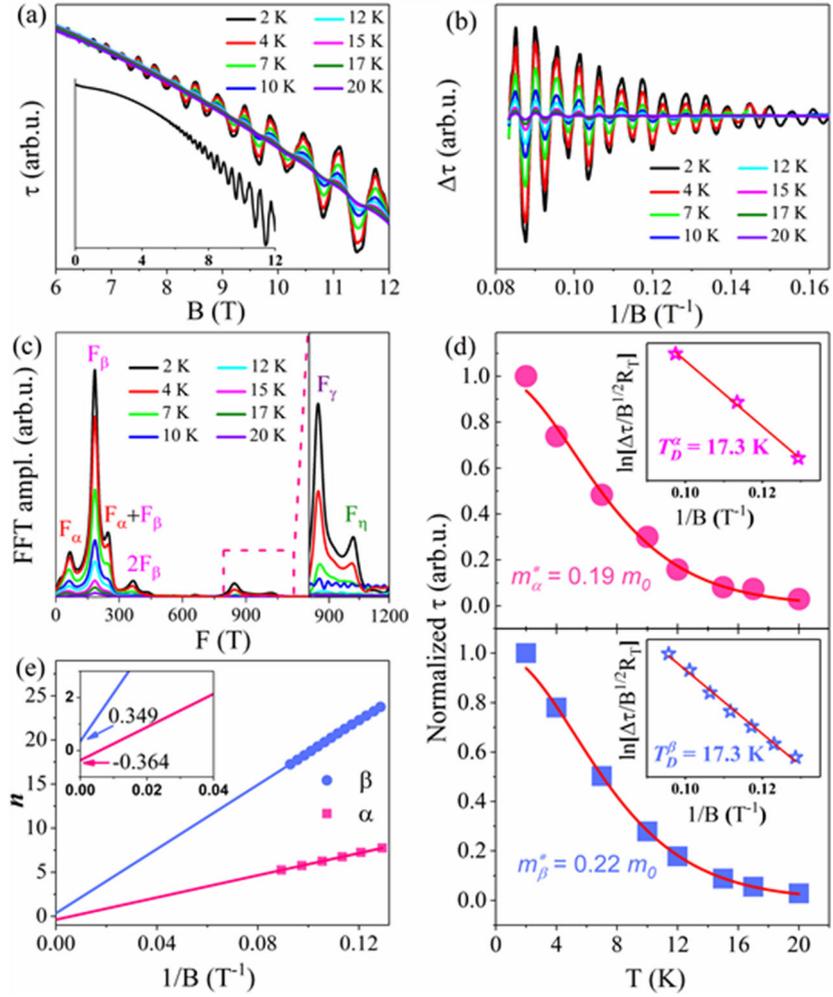

FIG. 2. (Color online) (a) Strong dHvA oscillations observed through magnetic torque $\tau$ measurements for $B$ along $b$-axis at various temperatures. Inset: the quantum oscillation at 2 K. (b) The oscillation component $\Delta\tau$ as function of $1/B$ at various temperatures. (c) FFT spectra of dHvA oscillations at different temperatures. Inset: the enlarged FFT spectra from 800 T to 1200 T. (d) The temperature dependence of the oscillation amplitudes for $\alpha$ and $\beta$ band at $1/B = 0.1135\ \text{T}^{-1}$ and $0.0956\ \text{T}^{-1}$, respectively. The small effect mass can be extracted from the well fitted curve (red line) using LK formula. Inset: The Dingle plots of dHvA oscillations at 2 K. (e) LL index $n$ versus $1/B$ at 2 K with the magnetic field parallel to the $b$-axis. Inset: nonzero intercepts with the $n$ axis obtained by the linear fit.

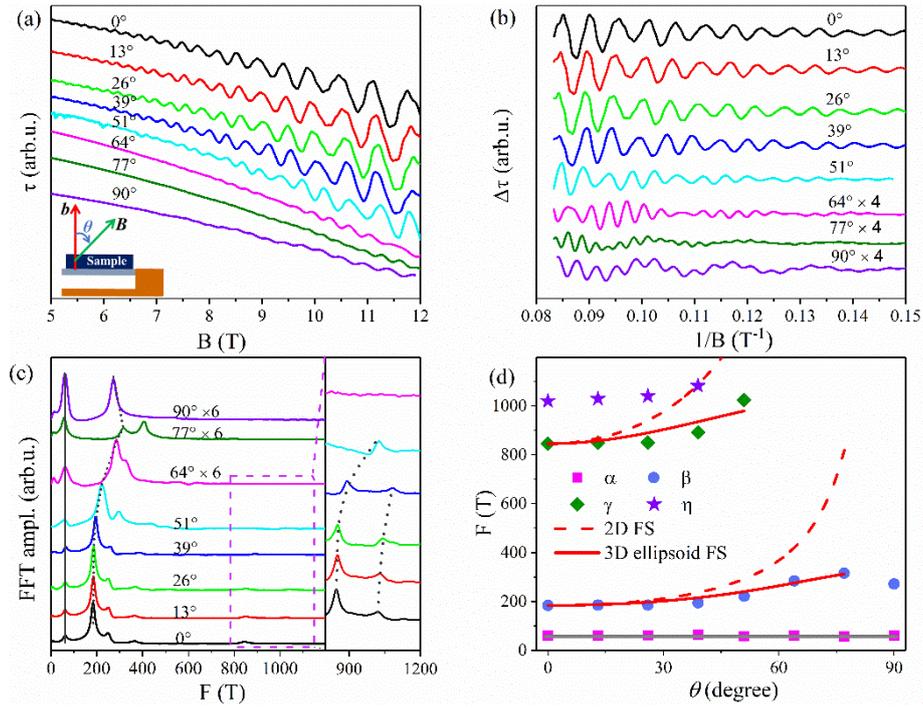

FIG. 3. (Color online) (a) The magnetic field dependent magnetic torque $\tau$ at various angles and $T = 2$ K. The results have been shifted for clarify. Inset: Schematic of the measurement configuration with $\theta$ the angle between the applied field and $b$-axis. (b) The oscillatory components after subtracting the smooth background as a function of $1/B$ at various angles. (c) The FFT spectra for different angles indicated. Inset: enlarged FFT spectra in the range of 800~1200 T. (d) The angular dependences of the oscillation frequencies of $\alpha$ (pink), $\beta$ (blue), $\gamma$ (olive), and $\eta$ (violet) bands. The dashed and solid red lines are the fitted curves using the standard 2D Fermi surface and 3D ellipsoid Fermi surface, respectively.

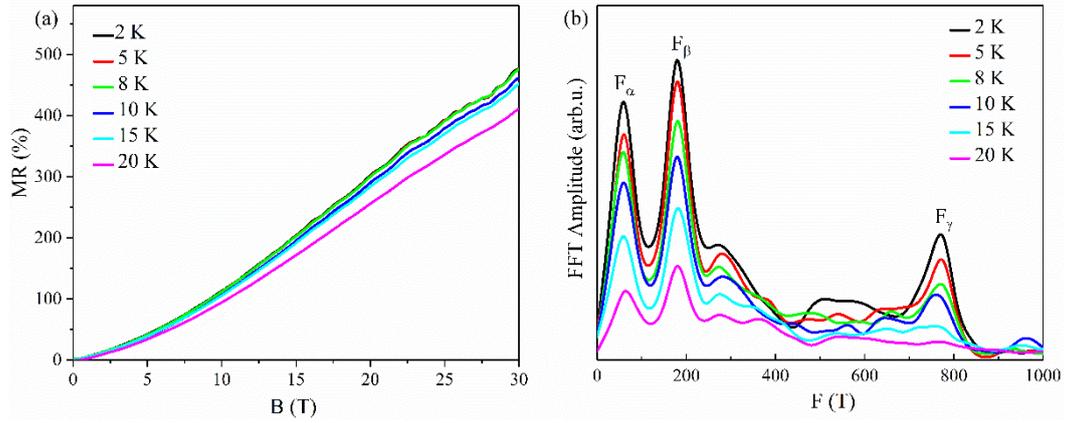

FIG. 4. (Color online) (a) The SdH oscillations of longitudinal MR at various temperatures with $B \parallel b$-axis. At high magnetic fields, the MR is linear without saturation. (b) FFT spectra of SdH oscillations at different temperatures.

Table. 1 The oscillation frequency, effective mass, Dingle temperature, quantum relaxation time, quantum mobility and the Berry phase obtained from dHvA oscillations in TaNiTe$_5$ single crystals.

| B//b | | F(T) | $m^*(m_0)$ | $T_D$(K) | $\tau_q$(10$^{-14}$s) | $\mu_q$ (cm$^2$V$^{-1}$s$^{-1}$) | $\Phi_B$ $\delta = 1/8$ | $\Phi_B$ $\delta = -1/8$ |
|---|---|---|---|---|---|---|---|---|
| | α | 63 | 0.19 | 17.3 | 7.03 | 649.4 | $-0.48\pi$ | $-0.98\pi$ |
| dHvA | β | 184 | 0.22 | 14.4 | 8.44 | 678.5 | $0.95\pi$ | $0.45\pi$ |
| | γ | 840 | | | | | | |
| | η | 1021 | | | | | | |